\documentclass[12pt]{iopart}

\usepackage{amssymb}
\usepackage{amsfonts}
\usepackage{amssymb}
\usepackage{graphicx}
\usepackage{fullpage}
\usepackage{color}
\usepackage{picture}
\usepackage{float}
\usepackage{fancyhdr}
\usepackage{footnote}
\usepackage{xcolor}
\usepackage{subfigure}
\usepackage{mathrsfs}
\usepackage{caption}

\begin{document}

\title{Molecular Dynamics Simulations for Anisotropic Thermal Conductivity of Borophene}

\author{Yue Jia $ ^1 $, Chun Li $ ^1  $  $ ^* $ , Jin-Wu Jiang $  ^2 $ , Ning Wei $ ^3 $ , Yang Chen $ ^3 $ and Yongjie Jessica Zhang $ ^4 $ }  
\address{$ ^1 $School of Mechanics, Civil Engineering and Architecture, Northewestern Polytechnical University, 710129 Xi'an, China}
%\address{$ ^1 $School of Mechanics, Civil Engineering and Architecture, Northewestern Polytechnical University, 710129 Xi'an, China}
\address{$ ^2  $    Shanghai Institute of Applied Mathematics and Mechanics, Shanghai University, 200433 Shanghai, China}
\address{$ ^3 $ College of Water Resources and Architectural Engineering, Northwest A$\&$F University, 712100 Yangling, China}
%\address{$ ^3 $  College of Water Resources and Architectural Engineering, Northwest A$\&$F University, 712100 Yangling, China}
\address{$ ^4 $ Department of Mechanical Engineering, Carnegie Mellon University, 15213 Pittsburgh, USA}

%\address{$ ^1 $ IOP Publishing, Temple Circus, Temple Way, Bristol BS1 6HG, UK}
\ead{lichun@nwpu.edu.cn (Corresponding Author)}
\vspace{10pt}
\begin{indented}
\item[]May 2017
\end{indented}

\begin{abstract}
The present work carries out molecular dynamics simulations to compute the thermal conductivity of the borophene nanoribbon  and the borophene nanotube  using the M$\ddot{\rm{u}}$ller-Plathe  approach. We investigate the thermal conductivity of the armchair and zigzag borophenes, and show the strong anisotropic thermal conductivity property of borophene. We compare the results of the borophene nanoribbon and the borophene nanotube, and find the thermal conductivity of the borophene is structure dependent. 

\end{abstract}

\noindent{\it Keywords}:  borophene, molecular dynamics, thermal conductivity
%
%
 %Uncomment for Submitted to journal title message

%\submitto{\NT}
\maketitle

\section{\label{sec:level1}Introduction} 

Boron sheet is commonly named as borophene, which consists of boron atoms with 2D nanostructure. In recent years, borophene has been synthesized through experiments in several labs \cite{Mannix2,Tai3,Feng4}, which makes a breakthrough for the borophene research. Borophene has attracted a lot of interests in research due to its fascinating physical and chemical properties. Because of its special anisotropic crystal structure, borophene also exhibits an optical anisotropy property, which can be used as transparent conductors in display technologies, photovoltics and flexible electronic devices \cite{Peng5}. Borophene has been found having intrinsic phonon-mediated superconductivity under estimated critical temperature when it is feasibly formed on a metal substrate \cite{Penev6}. Borophene can serve as an ideal electrode material due to its high capacity and chemical stability properties \cite{Zhang7}. Compared to other 2D materials,  borophene is  hard and brittle with negative Possion’s ratio \cite{Sun8,Wang9}. Borophene has been shown to have many anisotropic properties, such as special anisotropic crystal structure, highly anisotropic metallic behavior, large optical anisotropy. However, till now, there is few research conducted on the anisotropic thermal conductivity property of the borophene, which is important to realize related potential applications. 

In physics, thermal conductivity describes the property of a material to conduct heat. It is worth of noting that the existing research on the thermal conductivity of borophene is based on the first principle calculations \cite{Xiao14}. In particular, the thermal transport property of borophene is investigated through solving the Boltzmann transport equation based on the first principle calculations. Up to now, detailed study on the thermal conductivity of borophene based on molecular dynamics (MD)  is still lacking.

In this work, we apply the reverse nonequilibrium molecular dynamics (RNEMD) \cite{MP12} to investigate the thermal conductivity of borophene. Firstly, we analyze the thermal conductivity of two types of the borophene nanoribbon (BNR), namely the armchair borophene nanoribbons (ABNR) and the zigzag borophene nanoribbons (ZBNR), respectively. We calculate the thermal conductivity of BNR with different widths. Secondly, we generate the corresponding armchair borophene nanotubes (ABNT) and zigzag borophene nanotubes (ZBNT), and compute their thermal conductivity. We also compute the thermal conductivity of the borophene nanotube (BNT) with varying perimeters.  In addition, we also stretch the above model along the heat flux direction, and compute the corresponding thermal conductivities. In the end, we show borophene structures have clear thermal anisotropy property.

\section{Modeling and Computational Methods}

In this work, we perform MD simulations by using the Stillinger-Weber (SW) potential \cite{Jiang18} to describe the interaction in borophene. The nonlinear  SW potential is derived from the valence force field  model. The SW potential can be applied in the MD simulation of nonlinear physical or mechanical properties for borophene. All the MD simulations are achieved by the Large-scale Atomic/Molecular Massively Parallel Simulator (LAMMPS) software \cite{Plimpton19}, and the borophene structural visualization is obtained by the Visual Molecular Dynamics (VMD) \cite{webste11}.

\begin{figure}[H]
\begin{center}
\subfigure{
\includegraphics[scale=0.4]{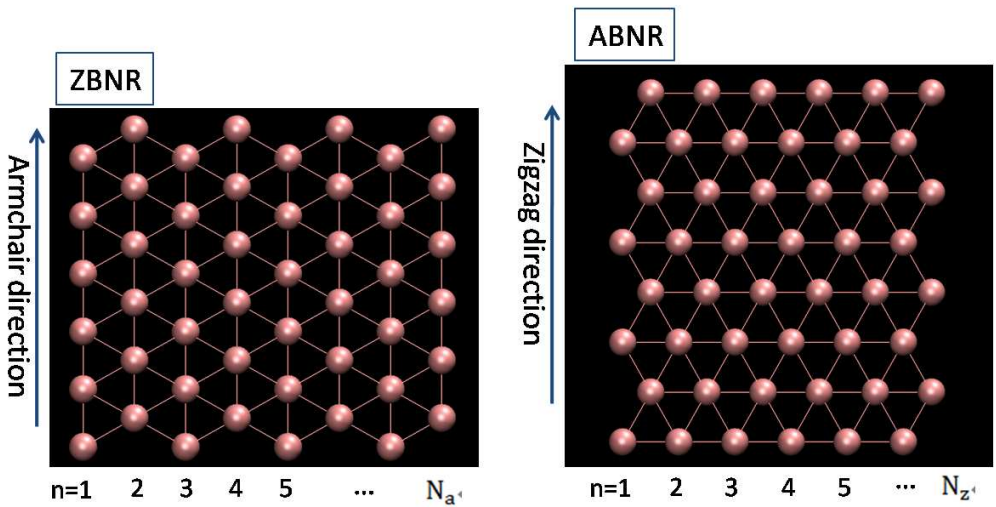}
}
\end{center}
\caption{ Atomic structure of typical armchair borophene nanoribbons  ($N_{a}$-ABNR) (left) and zigzag borophene nanoribbons ($N_z$-ZBNR) (right). The width of BNR is labeled by the number of dimer lines, $N_a$ for AGNR and $N_z$ for ZBNR, respectively.}\label{Figure1}
\end{figure}

Figure \ref{Figure1} shows the structure of borophene (top review), in which the structural parameters are from \cite{Zhou1}. The thermal conductivity is computed by the M$\ddot{\rm{u}}$ller-Plathe  (MP) approach \cite{MP12},  which is one kind of RNEMD. The MP method has been successfully applied for calculating the thermal conductivity of graphene \cite{Wei15,Wei16} and  phosphorene \cite{Zhang17}. Periodic boundary conditions are applied on the heat flux direction. The equilibrium temperature is given by 300$K$. After 300,000 steps, the system starts the heat flux processing. The MD simulation time step is 0.5 femtosecond (fs). The simulation processing is carried out around 8,000,000 steps in total. 
Each period borophene nanoribbon is divided into 50 slabs along the heat transfer direction. The first slab is set to be the cold region and the $26^{th}$ slab is set to be the hot region, which is shown in Figure \ref{Figure1}. The MP method is mainly to select the hottest atom (with the highest kinetic energy) in the cold region and the coldest atom (with the lowest kinetic energy), and pair these two atoms up and exchange their velocities. After repeating the procedure in a relatively long time, the system can reach the equilibrium state. The heat flux $ J $ is calculated by
\begin{equation}
J=\case{\sum_{N_{swap}}\case{1}{2}(mv_{h}^{2}-mv_{c}^{2})}{t_{swap}}
 \label{Equation5}
\end{equation}
where $t_{swap}$ and $ N_{swap} $ are the total swap time and the number of swaps, respectively. $ m $ is the atomic mass. $ v_{h} $ and $v_{c}$ stand for the velocities of the hottest and the coldest atoms in each time swap, respectively. The momentum exchanging is carried out every 5 fs under an NVE ensemble (constant energy and constant volume).  The nonequilibrium steady state can be achieved after 100 picosecond (ps) of the exchanging process. 

\begin{figure}[H]
\subfigure{
\includegraphics[scale=0.27]{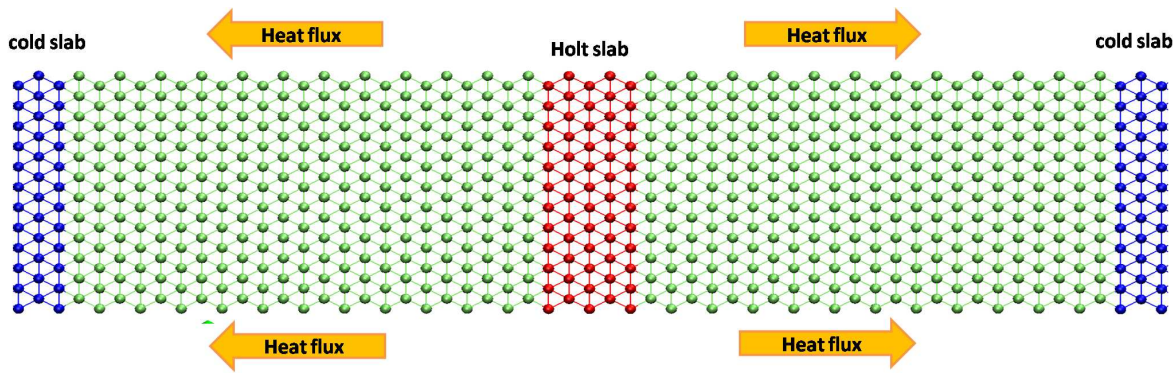}
}
\caption{ Schematics of the reverse non-equilibrium MD simulation. }\label{Figure2}
\end{figure}

\begin{figure}[H]
\begin{center}
\includegraphics[scale=0.26]{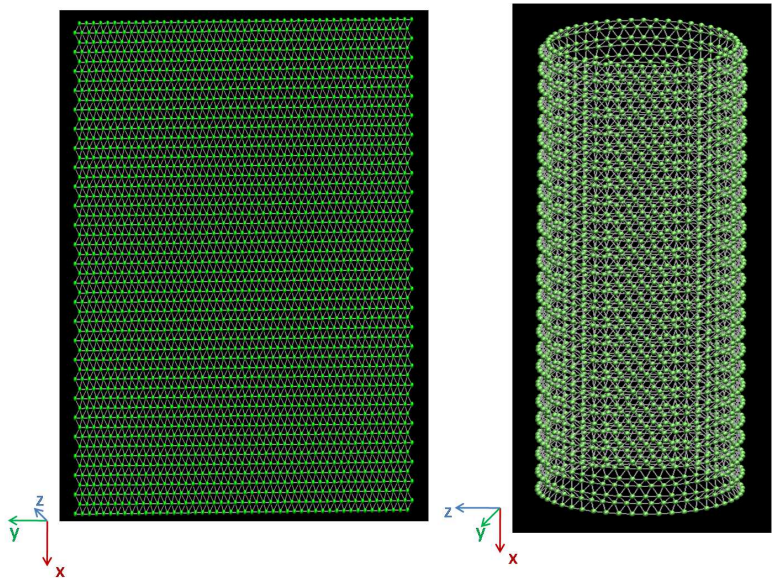}
\end{center}
\caption{ The configurations of the borophene nanoribbon and the boronphene nanotube. }\label{Figure3}
\end{figure}

\begin{figure}[H]
\begin{center}
\subfigure[]{
\includegraphics[scale=0.45]{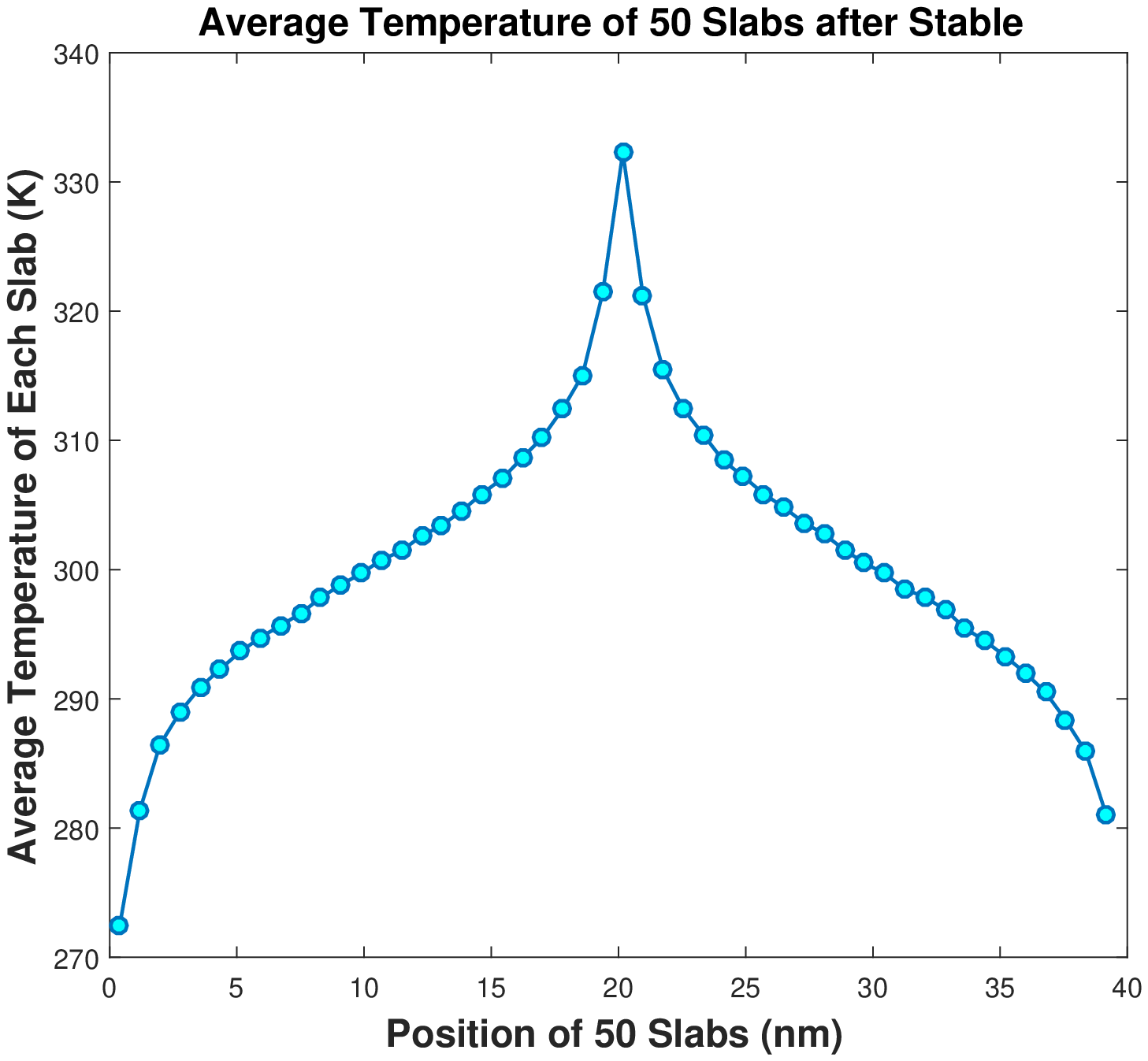}
}
\subfigure[]{
\includegraphics[scale=0.45]{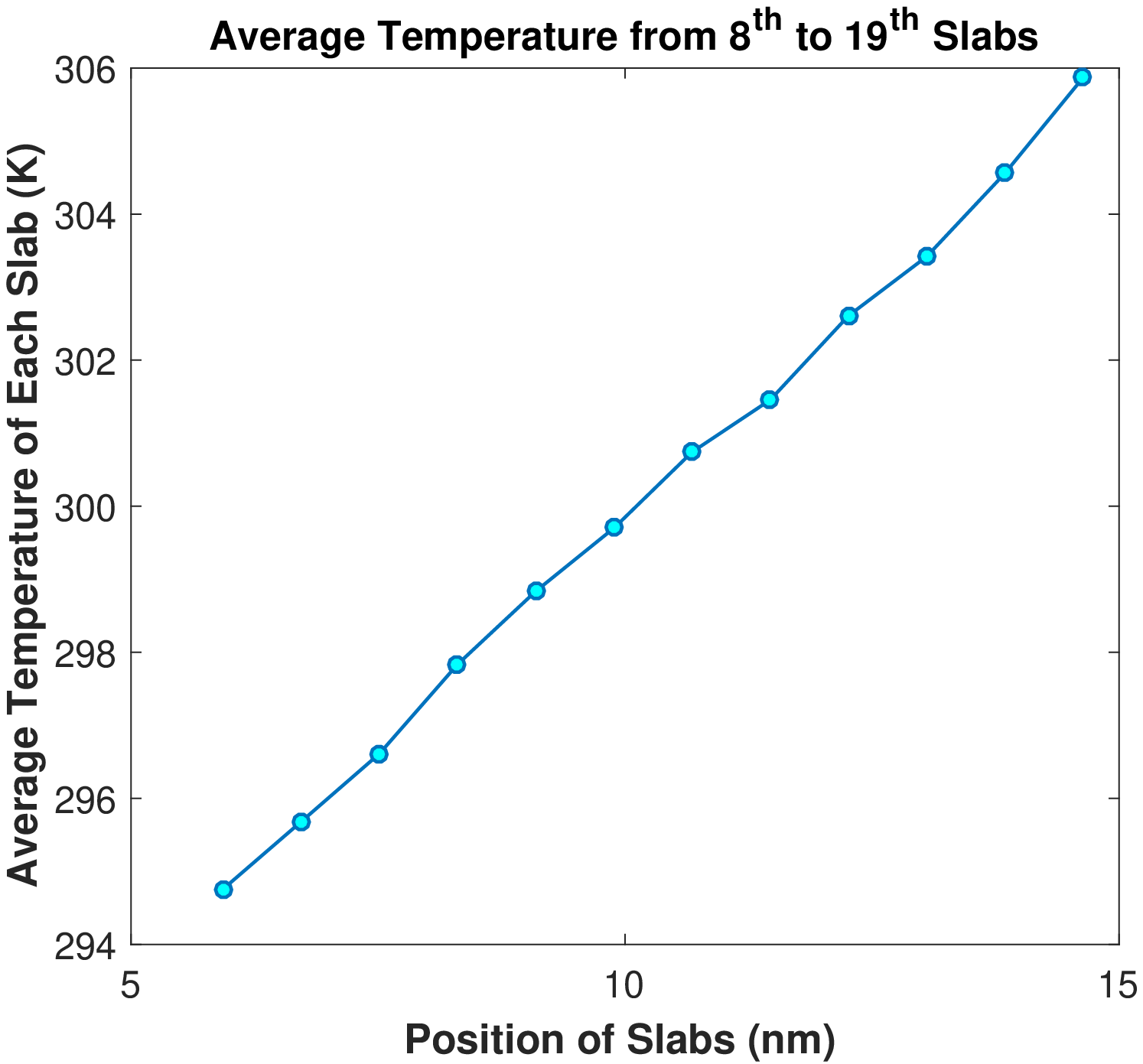}
}
\caption{ A typical example of temperature distribution. The temperature gradient is the function of the atom position along the $x$-direction which is parallel to the flux flow direction.  (a) averages the temperature of each slab. (b) takes the linear segment from the temperature profile and then obtain the gradient of the segment. }\label{Figure5}
\end{center} 
\end{figure}

\begin{figure}[H]
\begin{center}
\includegraphics[scale=0.2]{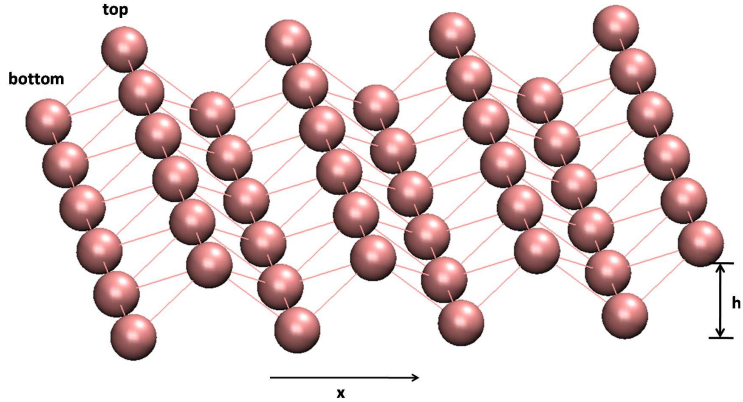}
\end{center}
\caption{ Perspective view illustrates the puckered configuration, with $h$ as the distance between the top and bottom chains along the out-of-plane $z$-direction. The pucker is perpendicular to the $x$-axis and parallel with the $y$-axis. }\label{Figure4}
\end{figure}

The temperature distribution is collected after 10 ns (with time step 0.5 fs) and the non-equilibrium steady state is reached. The temperature of each slab is calculated from
\begin{equation}
T_{i}(slab)=\case{2}{3Nk_{B}}\sum_{j}\case{p_{j}^{2}}{2m}
\end{equation}
where $T_{i}(slab)$ stands for the temperature of the $i^{th}$ slab. $ N $ is the number of the boron atoms in this slab. $k_{B}$ represents the Boltzmann’s constant and $p_{j}$ is the momentum of atom $j$. The temperature profiles are obtained by averaging results of around 60,000,000 steps which are collected every 30,000 steps. We show two typical examples of temperature distribution of  borophene in Figure \ref{Figure5}. The temperature gradient, $\case{dT}{dx}$, can be obtained by linear fitting in the labeled region as shown in Figure \ref{Figure5}b. The thermal conductivity is obtained using the Fourier law
\begin{equation}
\kappa = \case{J}{2A\partial T/\partial x}
\end{equation}
where $ A $ is the cross section area of the heat transfer, defined by the width multiply the thickness of the BNR. The thickness of borophene is 
assumed to be 0.911$\r{A}$  \cite{Zhou1}, which is the distance between the top chain and the bottom chain along the out-of-plane $z$-direction as shown in Figure \ref{Figure4}.

\section{ Results and Discussion}
\subsection{ Thermal Conductivity of Borophene Nanoribbons}
\begin{figure}[H]
\begin{center}
\includegraphics[scale=0.47]{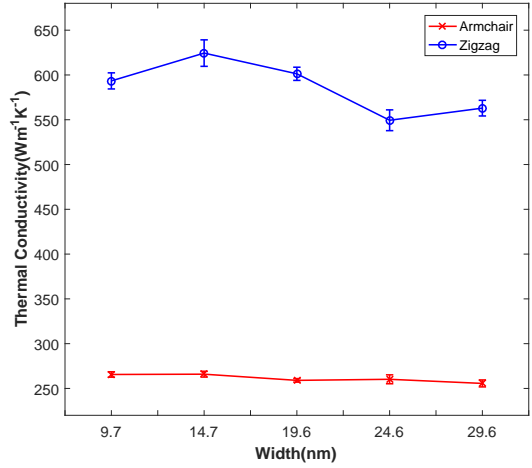}
\end{center}
\caption{ The calculated thermal conductivities of BNR with heat flux along the armchair direction (bottom line) and zigzag direction (above line). The length of the armchair cases is $ 39.551 \rm{nm} $, and the widths of the armchair cases are $ 9.684 \rm{nm} $, $ 14.526 \rm{nm} $, $ 19.852 \rm{nm} $, $ 24.694 \rm{nm} $ and $ 29.536 \rm{nm} $. The length of the zigzag cases is $ 39.704 \rm{nm} $, and the widths of the  zigzag cases are $ 9.744 \rm{nm} $, $ 14.903 \rm{nm} $, $ 19.489 \rm{nm} $, $ 24.648 \rm{nm} $ and $ 29.806 \rm{nm} $. }\label{Figure6}
\end{figure}

\begin{figure}[H]
\begin{center}
\includegraphics[scale=0.47]{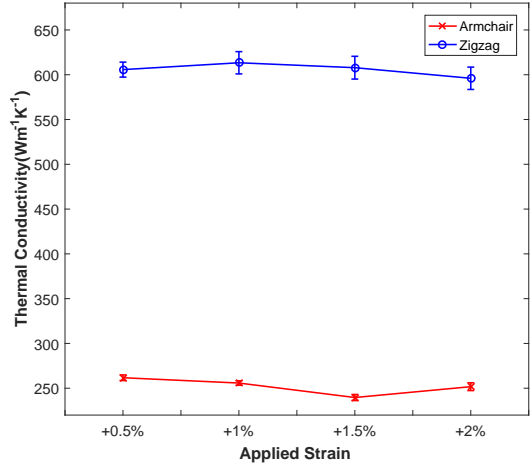}
\end{center}
\caption{ The calculated thermal conductivties of armchair direction BNR (bottom line) and zigzag direction BNR (above line) after applyed strains in heat flux direction. The applyed strains are $ +0.5 \% $, $ +1 \% $, $ +1.5 \% $ and $ +2 \% $, respectivily. The width of the armchair cases is $ 19.852 \rm{nm} $, and the lengthes of the cases are $ 39.947 \rm{nm} $, $ 40.346 \rm{nm} $, $ 40.746 \rm{nm} $ and $ 41.149 \rm{nm} $.  The width of the zigzag cases is $ 19.489 \rm{nm} $, and the lengths of the cases are $ 40.102 \rm{nm} $, $ 40.502 \rm{nm} $, $ 40.904 \rm{nm} $ and $ 41.308 \rm{nm} $. }\label{Figure7}
\end{figure}

We first performed MD simulations to study the thermal conductivity along the armchair and zigzag directions of BNR. Simulation samples are generated with the width varying from around 10 to 30 nm, and with the same length of 40 nm. We have tested the effect of the sample width on the simulation results, and  found that the thermal conductivity is width independent. From the Figure \ref{Figure6}, the thermal conductivity along the armchair and zigzag directions is found to be around 257.62$Wm^{-1} K^{-1}$  and 586.12$Wm^{-1} K^{-1}$, respectively. The borophene structures show a strong anisotropy property of the thermal conductivity. We also calculated the thermal conductivity of the BNR after applying certain among of strain (varying from 0.5$\% $ to 2$\%$) along the heat flux direction, and the numerical results are plot in Figure \ref{Figure7}. However, as the strain applying of the borophene increases, the corresponding thermal conductivity does not change. 

\subsection{  Thermal Conductivity of Borophene Nanotubes}
\begin{figure}[H]
\begin{center}
\includegraphics[scale=0.47]{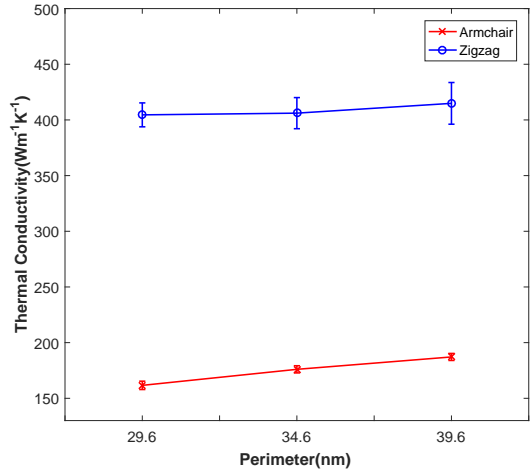}
\end{center}
\caption{ The calculated thermal conductivities of BNT with heat flux along the armchair direction (bottom line) and zigzag direction (above line). The length of the armchair cases is $ 39.551 \rm{nm} $, and the perimeters of the armchair cases are $ 29.852 \rm{nm} $, $ 34.694 \rm{nm} $,  and $ 39.536 \rm{nm} $. The length of the zigzag cases is $ 39.704 \rm{nm} $, and the perimeters of the  zigzag cases are $ 29.489 \rm{nm} $, $ 34.648 \rm{nm} $,  and $ 39.806 \rm{nm} $. }\label{Figure8}
\end{figure}

\begin{figure}[H]
\begin{center}
\includegraphics[scale=0.47]{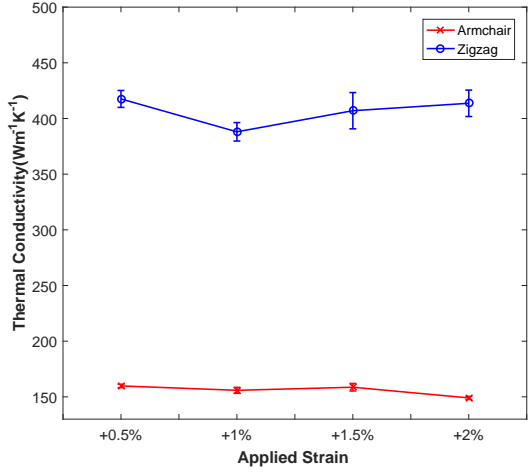}
\end{center}
\caption{ The calculated thermal conductivties of armchair direction BNT (bottom line) and zigzag direction BNR (above line) after applyed strains in heat flux direction. The applyed strains are $ +0.5 \% $, $ +1 \% $, $ +1.5 \% $ and $ +2 \% $, respectivily. The perimeter of the armchair cases is $ 29.852 \rm{nm} $, and the lengthes of the cases are $ 39.947 \rm{nm} $, $ 40.346 \rm{nm} $, $ 40.746 \rm{nm} $ and $ 41.149 \rm{nm} $.  The perimeter of the zigzag cases is $ 29.489 \rm{nm} $, and the lengths of the cases are $ 40.102 \rm{nm} $, $ 40.502 \rm{nm} $, $ 40.904 \rm{nm} $ and $ 41.308 \rm{nm} $. }\label{Figure9}
\end{figure}

In order to further examine the strong thermal anisotropy property of the borophene, we generate BNT structure and apply MD to study the thermal conductivity.  The in-plane thermal conductivities of the borophene along both the armchair and zigzag directions with different perimeters are showed in Figure \ref{Figure8}, respectively. We can observe that the thermal conductivity is around 174.86 and 408.03 $ Wm^{-1} K^{-1} $ in the armchair and zigzag directions, respectively. It can be seen that the thermal conductivity of the borophene is insensitive to their perimeters. However, the comparison results show strong thermal anisotropy phenomenon for BNT as well. In order to study the strains effect of BNT, we perform MD simulations on the in-surface thermal conductivity of BNT with different strains along the heat flux direction varying from  39.947 to 41.149 $ \rm{nm} $ for the armchair type and from 40.102 to 41.308 $ \rm{nm} $ for the zigzag type. From the calculated results in Figure \ref{Figure9}, we can observe that the thermal conductivity of BNT is not influenced by the strain.

\textbf{Discussion}. The above calculated numerical results show that borophene has strong anisotropy property for the thermal conductivity. We calculated the thermal conductivity of BNR with different widths (Figure \ref{Figure6}) and BNR with different perimeters (Figure \ref{Figure8}), and found that different widths and perimeters do not affect the thermal conductivity of the borophene. The thermal conductivity is not affected either when we stretch the BNR (Figure \ref{Figure7}) and BNT (Figure \ref{Figure9}) along the heat flux direction. If we compare the thermal conductivity of BNR (Figures \ref{Figure6} and \ref{Figure7}) with that of BNT (Figures \ref{Figure8} and \ref{Figure9}), the nanotube$\rm{s}^{'}$s is decreased.

\section{ Conclusion}
In this work, we use MD simulations to investigate the thermal conductivity of borophene. We calculate the thermal conductivity of BNR and BNT with the armchair and zigzag structures. We show the borophene structure has strong anisotropy property in the thermal conductivity. The thermal conductivity of the borophene with zigzag heat flux direction is much higher than that of the borophene with armchair heat flux direction. We vary the width of BNR and the perimeter of BNT, and find  that the thermal conductivity of borophene is not influenced by the different widths and perimeters. After comparing the numerical results of BNR and BNT, we notice the thermal conductivity of the borophene is structure dependent. The thermal conductivity of BNT is reduced compared to that of BNR. We also analyze the strain effect to the thermal conductivity of the borophene, which shows the thermal conductivity of borophene is insensitive to the applied strains.

\section*{Acknowledgements}

The authors would like to thank the support by the National Natural Science Foundation of China (Grant No. 11002109, 51210008, 11504225). Y. Zhang was supported in part by NSF CAREER Award OCI-1149591.

\section*{References}

\bibliographystyle{unsrt}

\end{document}